# Blockchain-oriented Software Engineering: Challenges and New Directions

Simone Porru, Andrea Pinna
University of Cagliari
Department of Electrical and Electronic Engineering
[simone.porru, a.pinna]@diee.unica.it

Michele Marchesi, Roberto Tonelli
University of Cagliari
Department of Informatics and Mathematics
marchesi@unica.it, roberto.tonelli@dsf.unica.it

*Abstract*—The Blockchain technology is reshaping finance, economics, and money to the extent that its disruptive power is being compared to that of the Internet and the Web in their early days. As a result, all the software development revolving around the Blockchain technology is growing at a staggering rate. In this paper, we acknowledge the need for software engineers to devise specialized tools and techniques for blockchain-oriented software development. From current challenges concerning the definition of new professional roles, demanding testing activities, and novel tools for software architecture, we take a step forward by proposing new directions on the basis of a curated corpus of blockchain-oriented software repositories, detected by exploiting the information enclosed in the 2016 Moody's Blockchain Report and the market capitalization of cryptocurrencies. Ensuring effective testing activities, enhancing collaboration in large teams, and facilitating the development of smart contracts all appear as key factors in the future of blockchain-oriented software development.

*Keywords*-blockchain; software engineering; smart contracts; cryptocurrencies;

## I. INTRODUCTION

In the past years, a lot of attention has been paid to the emerging concepts of Blockchain and Smart Contract. Organizations such as banking and financial institutions, and public and regulatory bodies, started to explicitly talk of the importance of these new technologies. Some observers are even talking of the dawn of a new era, declaring that *"we should think about the blockchain as another class of thing like the Internet [...]"* [1] and that the *"wide adoption of blockchain technology has the potential of reshaping the current financial services technical infrastructure."* [2].

From an economic and financial perspective, the overall capitalization of digital currencies is about 12 billions USD, as of October 2016, 80% of which is the capitalization of Bitcoins alone[1]. Venture capital investments in blockchain startups has been steadily increasing, from $93.8 million in 2013, to $315 million in 2014, to $490 million in 2015. Some people are even saying that Bitcoin and blockchain remind them of the Internet circa 1993, when huge amounts of venture capital started to flow into Internet startups, eventually leading to the emergence of companies such as Cisco, Yahoo, Google, Amazon and others.

Alongside these good news, there are also bad news, however. Ever since digital currencies started to represent a real monetary value, also hacks and attacks started. The "exchanges", Web sites allowing to store digital currencies and to trade them against other currencies, legal or digital, experimented big attacks. The biggest was the MtGox attack that occurred at the beginning of 2014, leading to a declared loss of $600 million. The latest was the Bitfinex attack of August 2016, with a loss of $65 million. Another remarkable exploit was that sustained by the DAO organization in June 2016, leading to a withdrawal of Ether digital currency funds worth $50-60 million. The Ethereum community defended themselves from this attack by performing a hard fork of the Ethereum software that forcibly recovered the stolen Ethers and gave them back to the original owners. Almost all of these attacks can be attributed to poor software development practices.

The feeling many software engineers have about such increased interest in blockchain technologies and, in particular, on the numerous software projects rapidly born and quickly developed around the various blockchain implementations is that of unruled and hurried software development. The scenario is that of a sort of competition on a first-come-first-served base which does not assure neither software quality, nor that all the basic concepts of software engineering are taken into account.

This paper aims at revealing the current issues and new directions for blockchain-oriented software engineering, and investigating the need for novel specialized software engineering practices for the blockchain sector. To this purpose, we:

1) identified the most relevant challenges for the state-of-practice blockchain-oriented software engineering;
2) highlighted peculiarities of some of the most popular blockchain-oriented software projects going on in the world;
3) proposed new research directions for blockchain-oriented software engineering, based on the results obtained from the previous steps.

The paper is organized as follows. In section II we report the most relevant related work; in section III we present the current challenges for blockchain-oriented software engineering; in section IV we provide statistical data on open source blockchain-oriented GitHub repositories, whose selection has been guided by the 2016 Moody's report on the blockchain sector [3]; in section V we propose new research directions on the basis of the dataset analysis; and finally, in section VI we report the conclusion.

## II. RELATED WORK

In [1], Swan discusses the applications of the Blockchain starting from the concept of cryptocurrency to that of the more advanced Smart Contract, in an attempt to demonstrate that the Blockchain can prove to be as disruptive a computing paradigm as the Internet and the mobile. When compared to the Blockchain, however, mobile development has raised less pressing issues from the software engineering perspective. In fact, the need for mobile-specific software engineering techniques has grown only recently, precisely as mobile applications started to be employed for business-critical purposes and, therefore, to require enhanced security and higher quality [4]. Conversely, business is the main reason behind the creation of blockchain-oriented software. The demand for security in blockchain applications is, therefore, even more pressing, and thus that for specialized software engineering processes.

---
[1] https://coinmarketcap.com/

III. BLOCKCHAIN-ORIENTED SOFTWARE ENGINEERING : CHALLENGES

We define as Blockchain-oriented Software (BOS) all software working with an implementation of a Blockchain. A Blockchain is a data structure characterized by the following key elements:

- data redundancy (each node has a copy of the Blockchain);
- check of transaction requirements before validation;
- recording of transactions in sequentially ordered blocks, whose creation is ruled by a consensus algorithm;
- transactions based on public-key cryptography;
- possibly, a transaction scripting language.

Considering the distinctive marks of a Blockchain, software engineers could benefit from the application of BOS-specific software engineering practices. Such practices would constitute the base of a Blockchain-oriented Software Engineering (BOSE).

In this section, we identify the most relevant BOSE challenges, and the issues which originate from them. To this purpose, for most challenges, we also provide excerpts from the SWEBOK[2] [5] to properly frame the related issues.

**New professional roles**. *"A recognized profession entails specialized skill development and continuing professional education"*[2]. Due to the business-critical nature of the Blockchain, finance and legal subjects have shown increasing interest toward BOS. At the same time, bootcamps for Blockchain developers are flourishing. The Blockchain sector will need professional figures with a well-defined skills portfolio comprising finance, law, and technology expertise. An example of a new role could be that of an intermediary between business-focused contractors with low technology expertise and IT professionals.

**Security and reliability**. *"Software Security Guidelines span every phase of the software development lifecycle"* and *"Software Reliability Engineered Testing is a testing method encompassing the whole development process"*[2]. A Blockchain must guarantee data integrity and uniqueness to ensure Blockchain-based systems are trustworthy. Ensuring security and reliability in BOS development might require specific methodologies such as Cleanroom Software Engineering [6] or thorough software reviews. Furthermore, mathematically sound analysis techniques could help enforcing reliability and security-related properties in blockchain-oriented applications.

[2]SWEBOK 2004 version

Testing techniques can also enhance system security and reliability. IBM recently expressed the need for continuous testing techniques to ensure blockchain software quality [3].

Testing techniques should be based on the nature of the application[2] which, in the case of BOS, is that of security-critical systems. In particular, there is a need for testing suites for BOS. These suites should include:

- Smart Contract Testing (SCT), namely specific tests for checking that smart contracts i) satisfy the contractors' specifications, ii) comply with the laws of the legal systems involved, and iii) do not include unfair contract terms.
- Blockchain Transaction Testing (BTT), such as tests against double spending and to ensure status integrity (e.g. UTXO[4]).

**Software architecture**. Specific design notations, macroarchitecture patterns, or meta-models may be defined for BOS development. To this purpose, software engineers should define criteria for selecting the most appropriate blockchain implementation, evaluating the adoption of sidechain technology, or the implementation of an ad-hoc blockchain. For example, Ethereum[5] has adopted a key-value store, which is a very simplistic database. By adopting a higher level data representation such as an Object Graph, it would be possible to speed up many operations which would otherwise be expensive using a key-value store [7].

**Modeling languages**. Blockchain-oriented systems may require specialized graphic models for representation. More specifically, existing models might also be adapted to BOS. UML diagrams might be modified or even created anew to account for the BOS specificities. For example, diagrams such as the Use Case Diagram, Activity Diagram, and State Diagram could not effectively represent the BOS environment.

**Metrics**. BOSE may benefit from the introduction of specific metrics. To this purpose, it could be useful to refer to the Goal/Question/Metric (GQM) method, that was originally intended for establishing measurement activities, but it can also be used to guide analysis and improvement of software processes [5].

Due to the distributed nature of the Blockchain, specific metrics are required to measure complexity, communication capability, resource consumption (e.g. the so-called gas in the Ethereum system), and overall performance of BOS systems.

IV. BLOCKCHAIN-ORIENTED SOFTWARE REPOSITORIES

In order to define new research directions for the BOSE on the basis of the state-of-practice of blockchain-oriented software, we conducted an exploratory study on a corpus comprising 1184 GitHub software repositories, which were identified with the use of the Moody's Blockchain Report [3] and CoinMarketCap [6]. We first identified the most relevant projects and players in the blockchain sector; then, we collected metadata on the corresponding BOS repositories. Information from the corresponding issue tracking systems were also considered.[7] In the remainder of this section, we provide details about the methodology used to build the BOS corpus, and the preliminary results we obtained.

[3]https://twitter.com/ibmsoftware/status/776605297037172736
[4]Unspent Transaction Output
[5]https://www.ethereum.org/

*A. Building a Dataset of Blockchain-oriented Software*

In this paper, we define a BOS project as a software project which contributes to the realization of a blockchain project. This definition includes both blockchain platforms, such as Bitcoin and Ethereum, and general blockchain software [8].

To identify BOS repositories we start from the corresponding blockchain projects. Moody's Investor Services recently identified more than 120 publicly announced blockchain projects in an in-depth report of the blockchain sector [3]. The projects list covers rated issuers across financial institutions, nonfinancial corporates and the official sector, and can be considered as a comprehensive list of blockchain projects going on in the world. The Moody's list is not a list of software projects; nevertheless, BOS projects stem from the blockchain projects on the list.

In addition to the Moody's list, we searched for the software associated to the currencies and assets with the highest capitalization, as reported by CoinMarketCap. Since we took the Moody's list as a baseline, we did not include currencies and assets with a lower capitalization than Stellar, the least capitalized cryptocurrency in the Moody's list for which we found a related software repository. Being Stellar on the 17th position at CoinMarketCap, we focused on the first 17 most capitalized currencies and assets.

Finding the software corresponding to a project in the Moody's list is not straightforward. When the project name is within a list entry (e.g., The Hyperledger Project), the software can be easily found by searching for the specified project name. When not specified, we searched for the involved blockchain startups, which often choose to publish their software on code-hosting platforms (e.g. GitHub, Bitbucket). As for the currencies and assets found on CoinMarketCap, this process was easier since each list entry is linked to the official website.

We focused on freely accessible, open-source software hosted on GitHub, a platform hosting the vast majority of the detected blockchain-oriented software projects. We decided to only consider software hosted on GitHub repositories because GitHub provides homogeneous metadata, which, in turn, allow us to compare projects on the basis of standard features. For instance, it is possible to evaluate project popularity by relying on the amount of stars given to a project by GitHub users, or on the number of forks stemming from it.

[6]https://coinmarketcap.com/. We refer to the market capitalization of September 23, 2016
[7]We focused on Github-hosted projects (see section IV-A), which come with the integrated GitHub issue tracking system

*B. Dataset Analysis*

At the end of the selection process, we identified 52 GitHub accounts, which comprise 1184 repositories. We extracted information on popularity (*Stargazers*), programming languages, community involvement (*Contributors*, *Open Issues*, *Watchers*, *Forks*), and age (time elapsed since creation).

To focus on the most relevant repositories, we only considered those that i) are base repositories (not a fork from a previously existing repository), ii) had been updated in the previous 30 days, and iii) were created more than 30 days before (i.e. have been modified at least once since creation) [8]. By using these criteria, we retained 193 repositories out of the initial 1184.

*C. Preliminary Results*

Figure 1 reports the most used programming languages among the 193 retained repositories. JavaScript, Python, Go, C++, and Ruby are the top 5 languages, with BOS JavaScript repositories accounting for more than 30% of the total. It is interesting to note the presence of Python and Go in the podium, especially in comparison with the number of Java repositories–not reaching 4% of the total.

Table I shows that among the top 10 most popular repositories (i.e. those with the highest number of *Stargazers*) `ethereum/mist` and `coinbase/toshi` were created less than 1 year and roughly 2 years ago respectively. As expected, Bitcoin is the most popular project, neatly distinguished as for stargazers (9966) and contributors (396). Ethereum is also very popular, with three associated repositories in the top 10; in particular, the one written in Go is just behind the main Bitcoin repository. The top 10 BOS repositories were created around 4 years ago on average, and most of them have a considerable number of open issues. The statistics about forks are staggering, topping at 4266 for the main bitcoin repository, followed by the Go repository from Ethereum with 695 forks.

## V. BLOCKCHAIN-ORIENTED SOFTWARE ENGINEERING: NEW RESEARCH DIRECTIONS

Based on the results from section IV, we hereby suggest some new research directions for BOSE.

**Testing.** A recent study on over 50000 GitHub projects [9] has proved that a bigger team size leads to a higher number of test cases, whereas the number of test cases per developer decreases with an increase in the team size. It would be interesting to investigate whether the same can be said about BOS, considering that the most popular repositories have an unusually high number of contributors, even for open-source projects. For instance, almost 400 GitHub users are contributing to the `bitcoin/bitcoin` repository, as reported in Table I.

**Collaboration.** The high number of voluntary contributors testifies to the attractiveness of BOS in the open source landscape. A large base of voluntary contributing members has been shown to be a pivotal success factor in OSS evolution [10]. To achieve sustainable development and improve software quality, specific practices to enhance the synergy between the system and the community would be highly beneficial to BOSE [11].

[8]All data were retrieved on September 23, 2016

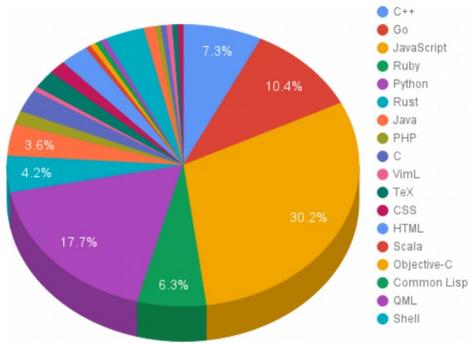

Fig. 1. Languages across 193 repositories

TABLE I
EXTRACTED STATISTICS ACROSS THE TOP 10 BOSR EPOSITORIES

| GitHub Repository | stargazers | contributors | open issues | age (days) | watchers | forks | first language |
|---|---|---|---|---|---|---|---|
| bitcoin/bitcoin | 9966 | 396 | 547 | 2105 | 1211 | 4266 | C++ |
| ethereum/go-ethereum | 2160 | 78 | 285 | 1002 | 367 | 695 | Go |
| ledger/ledger | 1813 | 108 | 14 | 3055 | 103 | 255 | C++ |
| digitalbazaar/forge | 1584 | 41 | 137 | 2260 | 103 | 241 | JavaScript |
| ripple/ripple-client | 1244 | 51 | 21 | 1437 | 968 | 486 | JavaScript |
| ethereum/mist | 1168 | 35 | 198 | 471 | 210 | 299 | JavaScript |
| dogecoin/dogecoin | 1153 | 300 | 52 | 1022 | 149 | 505 | C++ |
| ripple/rippled | 1144 | 53 | 118 | 1782 | 246 | 338 | C++ |
| coinbase/toshi | 839 | 18 | 97 | 749 | 98 | 187 | Ruby |
| ethereum/cpp-ethereum | 723 | 89 | 212 | 1001 | 196 | 270 | C++ |

**Enhancement of testing and debugging for specific programming languages.** Figure 1 shows that a number of programming languages such as Go, Python, and Ruby are gaining increasing popularity among BOS projects. This arises the need for enhanced testing and debugging suites, tailored upon the most popular BOS languages. Indeed, Java testing suites have undergone much more testing than Go.

In addition, as BOS projects work with the Blockchain, which is distributed by definition, testing in isolation would require properly mocking objects capable of effectively simulate the Blockchain.

**Creation of software tools for smart contract languages.** The implementation of Smart Contract Development Environments (SCDEs)–the blockchain-oriented declination of IDEs– might be pivotal for the building and diffusion of BOS expertise. Such environments could streamline smart contract creation through specialized languages (e.g. Solidity, a language designed for writing contracts in Ethereum).

## VI. CONCLUSION

Blockchain-oriented software engineering must respond to many challenges and define new directions to allow effective software development. In the present work, we highlighted the most evident issues of state-of-art Blockchain-oriented software development, by advocating the need for new professional roles, enhanced security and reliability, novel modeling languages, and specialized metrics. In addition, we used the 2016 Moody's Blockchain Report and the market capitalization of cryptocurrencies to build a dataset of blockchain-oriented software repositories. After retaining 193 repositories out of 1184, we extracted statistical information on popularity, collaboration, repository age, and programming languages. On the basis of the results of the analysis, we proposed new directions for blockchain-oriented software engineering, focusing on collaboration among large teams, testing activities, and specialized tools for the creation of smart contracts.

## REFERENCES

[1] M. Swan, *Blockchain: Blueprint for a new economy*. O'Reilly Media, Inc., 2015.
[2] Unicredit, "Blockchain technology and applications from a financial perspective," 2016.
[3] N. Caes, R. Williams, E. H. Duggar, and M. R. Porta, "Robust, cost-effective applications key to unlocking blockchain's potential credit benefits," 2016.
[4] A. I. Wasserman, "Software engineering issues for mobile application development," in *Proceedings of the FSE/SDP workshop on Future of software engineering research*. ACM, 2010, pp. 397–400.
[5] A. Abran, J. W. Moore, P. Bourque, R. Dupuis, and L. Tripp, "Swebok: Guide to the software engineering body of knowledge 2004 version," *IEEE Computer Society, Los Alamitos, California*, 2004.
[6] H. D. Mills, M. Dyer, and R. C. Linger, "Cleanroom software engineering," *IEEE Software*, vol. 4, no. 5, p. 19, 1987.
[7] D. Larimer. Introducing bitshares object graph. [Online]. Available: https://goo.gl/TWWSif. Accessed on 2016-10-20.
[8] G. Hileman, "State of blockchain q1 2016," 2016. [Online]. Available: http://www.coindesk.com/state-of-blockchain-q1-2016/
[9] P. S. Kochhar, T. F. Bissyandé, D. Lo, and L. Jiang, "Adoption of software testing in open source projects–a preliminary study on 50,000 projects," in *Software Maintenance and Reengineering (CSMR), 2013 17th European Conference on*. IEEE, 2013, pp. 353–356.
[10] K. Nakakoji, Y. Yamamoto, Y. Nishinaka, K. Kishida, and Y. Ye, "Evolution patterns of open-source software systems and communities," in *Proceedings of the international workshop on Principles of software evolution*. ACM, 2002, pp. 76–85.
[11] M. Aberdour, "Achieving quality in open-source software," *IEEE software*, vol. 24, no. 1, pp. 58–64, 2007.